\documentclass{mn2e}

\usepackage{graphicx,amsmath,amssymb,subfigure}

\begin{document}

\title[The Likelihood Ratio as a tool for Radio Continuum Surveys]{The Likelihood Ratio as a tool for Radio Continuum Surveys with SKA precursor telescopes \thanks{Based on observations collected at the European Organisation for Astronomical Research in the Southern Hemisphere, Chile, VIDEO: 179.A-2006}\thanks{ Based on observations obtained with MegaPrime/MegaCam, a joint project of CFHT and CEA/DAPNIA, at the Canada-France-Hawaii Telescope (CFHT) which is operated by the National Research Council (NRC) of Canada, the Institut National des Science de l'Univers of the Centre National de la Recherche Scientifique (CNRS) of France, and the University of Hawaii. This work is based in part on data products produced at TERAPIX and the Canadian Astronomy Data Centre as part of the Canada-France-Hawaii Telescope Legacy Survey, a collaborative project of NRC and CNRS.}}

\author[McAlpine et al.]
{K. McAlpine$^{1}$, D.J.B. Smith$^2$, M.J. Jarvis$^{2,3}$, D.G. Bonfield$^2$, S.Fleuren$^4$\\
\footnotesize
$^{1}$Department of Physics and Electronics, Rhodes University, Grahamstown,
6139, South Africa\\
$^{2}$Centre for Astrophysics, Science \& Technology Research Institute,
University of Hertfordshire, Hatfield, Herts, AL10 9AB, UK\\
$^{3}$Physics Department, University of the Western Cape, Cape Town, 7535, South Africa\\
$^{4}$School of Mathematical Sciences, Queen Mary, University of London, Mile End Road, London, E1 4NS, UK
}

\maketitle

\begin{abstract}
In this paper we investigate the performance of the likelihood ratio method as a tool for identifying optical and infrared counterparts to proposed radio continuum surveys with SKA precursor and pathfinder telescopes.  We present a comparison of the infrared counterparts identified by the likelihood ratio in the VISTA Deep Extragalactic Observations (VIDEO) survey to radio observations with 6, 10 and 15~arcsec resolution. We cross-match a deep radio catalogue consisting of radio sources with peak flux density $>$~60~$\mu$Jy with deep near-infrared data limited to $K_{\mathrm{s}}\lesssim$~22.6. Comparing the infrared counterparts from this procedure to those obtained when cross-matching a set of simulated lower resolution radio catalogues indicates that degrading the resolution from 6~arcsec to 10 and 15~arcsec decreases the completeness of the cross-matched catalogue by approximately 3 and 7~per~cent respectively. When matching against shallower infrared data, comparable to that achieved by the VISTA Hemisphere Survey, the fraction of radio sources with reliably identified counterparts drops from $\sim$89\%, at $K_{\mathrm{s}}\lesssim$22.6, to 47\% with $K_{\mathrm{s}}\lesssim$20.0. Decreasing the resolution at this shallower infrared limit does not result in any further decrease in the completeness produced by the likelihood ratio matching procedure. However, we note that radio continuum surveys with the MeerKAT and eventually the SKA, will require long baselines in order to ensure that the resulting maps are not limited by instrumental confusion noise.

\end{abstract}

\begin{keywords}

\end{keywords}

\section{Introduction}\label{sec:intro}

Future radio continuum surveys with SKA pathfinder instruments including the Evolutionary Map of the Universe (EMU; Norris et al. 2011),  Westerbork Observations of the Deep APERTIF Northern-Sky (WODAN) and the surveys to be conducted with the Low-Frequency Array (LOFAR; R\"ottgering et al. 2011) aim to map large areas of the  sky ($\gtrsim$ 1000 sq degrees) to very deep levels (5$\sigma \simeq$ 50 $\mu$Jy) with the purpose of addressing a number of key astrophysical questions. Specifically they aim to constrain the cosmic star-formation history of the Universe out to z$\sim$2, to map the evolution of active galactic nuclei (AGN) to the edge of the universe and to probe the influence of postulated AGN feedback mechanisms on star-formation activity and the formation and evolution of galaxies (Norris et al. 2011). The wide-field and source density of these surveys may be particularly suited to cosmological studies (e.g. Raccanelli et al. 2011).  On the other hand, smaller area surveys with the eMERLIN (Muxlow 2010), EVLA (e.g. Myers et al. 2010) and the future MeerKAT MIGHTEE surveys (Jarvis 2011) will push to much deeper flux densities ($\lesssim 5\mu$Jy rms) to obtain a clear census of activity in the Universe up to higher redshifts, traditionally thought of as the realm of rest-frame optical and ultra-violet surveys. As radio observations are unaffected by dust extinction they have the potential to provide a dust-unbiased probe of the accretion and star-formation history of the Universe at these redshifts. 

At flux densities greater than a few mJy radio surveys detect almost exclusively 'radio-loud' AGN whereas at lower flux density levels an increasing fraction of the radio source population is identified with star-forming galaxies. The relative fraction of AGN and star-forming galaxies present at these low flux densities is not well-determined  but observational studies and extrapolations of the radio luminosity functions of both populations indicate that AGN constitute a significant fraction  (up to 50\%) of the radio source population even at levels of ten's of $\mu$Jy (Sadler et al. 2002; Jarvis \& Rawlings 2004;  Simpson et al. 2006;  Seymour et al. 2008; Kellermann et al. 2008; Padovani et al. 2009; 2011). Thus there is no observational regime in which one can assume a faint radio source is simply associated with a starburst galaxy, and complementary datasets at optical and infrared wavelengths will play a vital role in classifying these faint radio sources. Furthermore, optical and near-infrared imaging are crucial for providing estimates of a radio source redshift through photometric redshift techniques or through follow-up spectroscopy at the position of the optical/near-infrared counterpart. 

Attaining the scientific goals of future deep continuum surveys is thus largely dependent on the ability to identify the correct multi-wavelength counterparts to the faint radio sources. Identifying the counterparts to a large fraction of higher redshift radio sources will require very deep complementary multiwavelength datasets. Given the resolution of these planned radio surveys, 10~arcsec for EMU, 15~arcsec for WODAN, it is possible that this necessary increase in depth of complementary datasets will complicate the cross-identification process. The resolution and signal to noise ratio of the radio observations determines a lower limit on the positional accuracy of the faint radio source positions dictating that the true counterparts might be located anywhere within this positional uncertainty. However the higher source density of the deeper complementary data creates an increased probability of both multiple counterparts and spurious alignments being detected within these large search radii, rendering nearest neighbour matching techniques unreliable.

A method which is often used to identify counterparts to low resolution radio observations is the Likelihood Ratio (LR) technique. This technique was first developed by Richter (1975) and later expanded upon by de Ruiter et al. (1977); Sutherland \& Saunders (1992) and Ciliegi et al.~(2003). The method combines information about the brightness distribution of the complementary higher resolution data and the positional errors in both the radio source catalogue and the complementary dataset to determine the most likely counterpart. It is thus of interest to determine how this more sophisticated matching technique will perform in the proposed case of low resolution ($\gtrsim$10 arcsec) radio observations matched to high resolution very deep infrared and optical catalogues.  
To investigate this question this paper will degrade the positional accuracy of a set of deep radio observations taken with 6 arcsec resolution and produce a number of simulated catalogues whose positional accuracies are consistent with observations taken at 10 arcsec and 15 arcsec resolution. We use a LR analysis, similar to that in Smith et al. (2011), to determine reliable near-infrared (NIR) counterparts to both the original catalogue and the simulated `low resolution' catalogues and compare the results.

 The paper is structured as follows: sections 2 and 3  outline the radio and infrared observations used in the matching procedure. Section 4 gives the details of the LR technique used for matching, section 5 explains the procedure used to simulate the low resolution catalogue and section 6 gives the results of the comparison. Section 7 summarises the effects of blended radio sources  and our conclusions are presented in section 8. All magnitudes are quoted in the AB magnitude system. We assume that $H_{0}$=70~km.s$^{-1}$.Mpc$^{-1}$ and a $\Omega_{M}$=0.3 and $\Omega_{\Lambda}$=0.7 cosmology throughout this paper.

\section{Radio Observations}
The radio survey used in this analysis consists of Very Large Array (VLA) observations at 1.4~GHz undertaken by Bondi et al. (2003). These observations were used to  produce a mosaic image with nearly uniform noise of $\sim$17~$\mu$Jy over 1 square degree centred at $\alpha$(J2000)=$\mathrm{2^h26^m00^s}$ and $\delta$(J2000)=$\mathrm{-4^{d}30' 00''}$. They were taken in the VLA B-configuration and have a FWHM synthesised beamwidth of approximately 6~arcsec.   
 
A catalogue of radio sources was extracted from the mosaiced image using the \textsc{aips} (Greisen 2003) Search and Destroy (SAD) task, retaining only sources with a peak flux to local noise ratio of $\gtrsim 5$. This procedure resulted in a catalogue of 1054 radio sources whose peak flux densities exceed 60~$\mu$Jy. Of these 1054 sources 19 are identified as multiple component radio sources. Radio sources are identified as multiple component sources if their individual components meet the following  three criteria, the components are separated by $<$~18~arcsec, have peak flux ratios $<$~3, and all components have a peak flux $>$~0.4~mJy/beam. Further details of the calibration, catalogue extraction and multi-component classification procedures are outlined in Bondi et al. (2003). 

For the purpose of evaluating the performance of the LR technique we disregard the multiple component radio sources in the catalogue as the LR relies on knowledge of the expected position of the infrared counterpart source and the associated errors on this position, and both of these quantities are poorly determined in the case of multiple component radio sources. We also exclude single component radio sources whose morphologies are asymmetric or elongated as the position of the potential counterpart source is not well known in these cases. These criteria result in the exclusion of a further 3 radio sources from our input catalogue. We acknowledge that such sources will be important components of future wide-area radio surveys, however they are not key to the work presented here.

\section{Infrared Observations}
\begin{figure}
 \includegraphics[width=1\columnwidth]{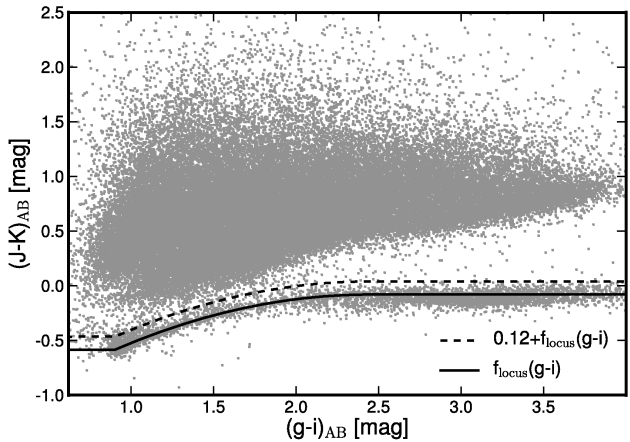}
\caption{Fit to the stellar locus used to remove stars from the infrared cross-matching catalogue. The solid line indicates our fit to the stellar locus in the combined VIDEO/CFHTLS-D1 dataset. The dashed line indicates the separation criteria applied and objects below the dashed line were removed from the cross-matching catalogue.}
\label{fig:sgsep}
\end{figure}
The square degree of VLA radio observations has been observed with the VISTA Deep Extragalactic Observations (VIDEO; Jarvis et al. in prep.) Survey. The VIDEO survey is a 12 sq. degree survey over three fields with the Visible and Infrared Survey Telescope for Astronomy (VISTA) and is designed to investigate the formation and evolution of galaxies and galaxy clusters. The survey provides photometry in the $Z$,$Y$,$J$,$H$, $K_{\mathrm{s}}$ bands to 5$\sigma$ depths of 25.7, 24.6, 24.5, 24.0, 23.5 magnitudes (2~arcsec diameter apertures) respectively. This field also coincides with the Canada-France-Hawaii Telescope Legacy Survey (CFHTLS) D1 field which provides additional photometry in the $u*$,$g'$,$r'$,$i'$,$z'$ optical bands.

To minimize the number of spurious faint sources in the combined catalogue used for cross-matching we retain only those sources with $K_{\rm s} < 22.6$ (Petrosian magnitude). We also disregard radio sources whose counterparts are affected by the presence of nearby saturated stars in the infrared images.

As infrared photometry to the depths of the VIDEO survey will not be available for several years across the 1000's of square degrees surveyed by future radio continuum surveys it is of interest to determine the likely completeness achieved when cross-matching these large radio surveys against infrared surveys with similarly large sky coverage. One of the largest near-infrared surveys currently being undertaken is the VISTA Hemisphere Survey (VHS) which is surveying the entire Southern Hemisphere to a 5$\sigma$ depth of $K_{s}$=20.0. The VHS thus represents one of the deepest complementary datasets covering the entire survey area of the wide-field EMU survey.  We investigate the completeness that cross-matching to this survey will produce by performing a LR cross-match to our VIDEO catalogue limited to detections with $K_{s}<$~20.0

\subsection{Star-galaxy separation}
To remove contaminating stars from the combined NIR/optical catalogue, which are unlikely to be genuinely associated with radio sources, we employ a colour based criteria similar to that used by Baldry et al. (2010) in their star-galaxy separation algorithm for target selection in the GAMA survey. To achieve this separation we fit the stellar locus in ($J$-$K$)$_{\mathrm{AB}}$ versus ($g$-$i$)$_{\mathrm{AB}}$ colour space with a quadratic $f_{\mathrm{locus}}(x)$ given by:

 \begin{align} &&&-0.58 & & & x<0.4 \notag \\
f_{\mathrm{locus}}(x)&=&&-0.88+0.82x-0.21x^2 & \mathrm{for}\hspace{1pt}&& 0.4<x<1.9 \notag \\
&&&-0.08&& & x>1.9 \notag \\
\end{align}

Objects are removed from our cross-matching catalogue if their colours meet the criteria:\begin{equation} \label{eqn:locus}
                                                                            J-K<0.12+f_{\mathrm{locus}}(g-i)
                                                                           \end{equation}
 Figure \ref{fig:sgsep} illustrates the fit to the stellar locus and the separation criteria used.

 Another method to achieve star-galaxy classification is to determine how well an object is resolved in the optical/NIR image, with stars and quasars being unresolved.  Object detection and photometry for the VIDEO survey was achieved using \textsc{sextractor} (Bertin \& Arnouts 1996); full details of the extraction will be given in Jarvis et al. (in prep.). This package  uses a neural network to assess how well resolved an object is and thereby determines a likelihood that the object is a star or galaxy. This likelihood is expressed as a CLASS\_STAR estimate between 0 and 1.0 with stars having measurements close to 1.0 and galaxies close to zero. An inspection of these   CLASS\_STAR estimates reveal that the total contribution of stars to the VIDEO catalogue after implementing our colour threshold is $<$5\%. We chose to use the criteria in equation \ref{eqn:locus} rather than a  straightforward CLASS\_STAR cut to avoid removing quasars, which are also unresolved in optical images and may be genuine counterparts to the radio sources. However it should be noted that the criteria in equation \ref{eqn:locus}  also removes a small number of quasars from our final NIR catalogue.

\section{Likelihood ratio}\label{sec:lr} 

The likelihood ratio is the ratio of the probability that a given source and counterpart are related to the probability that they are unrelated. It is given by the relationship  (Sutherland \& Saunders 1992): \begin{equation}
                                                                                                                                                                                                           \mathrm{LR}=\frac{q(m)f(r)}{n(m)}
                                                                                                                                                                                                          \end{equation} where $f(r)$ is the radial probability distribution function of the offsets between the radio and infrared positions, $q(m)$ is the expected distribution of the true infrared counterparts as a function of $K_\mathrm{s}$-band magnitude and $n(m)$ is the magnitude distribution of the full catalogue of $K_{\mathrm{s}}$-band detected objects.

The radial probability distribution $f(r)$ is given by a  Gaussian:\begin{equation}
                                                                            f(r)=\frac{1}{2\pi\sigma_{\mathrm{pos}}^{2}}\mathrm{exp}\left(\frac{r^2}{2\sigma_{\mathrm{pos}}^2}\right)
                                                                           \end{equation}
where $r$ is the offset between the radio and infrared position and $\sigma_{\mathrm{pos}}$ is the combined positional error of the radio and infrared sources.

\begin{figure}
 \includegraphics[width=1.0\columnwidth]{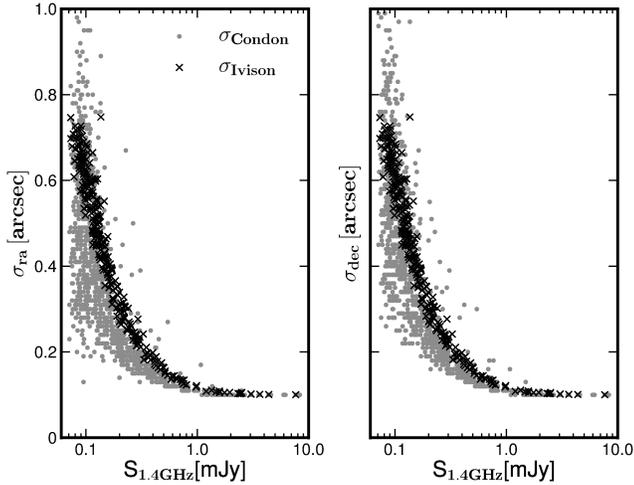}
\caption{Comparison between the errors in the radio source positions calculated using the relationships in Condon (1997) $\sigma_{\mathrm{Condon}}$ and  the error estimates used in our LR analysis $\sigma_{\mathrm{Ivison}}$ based on the relationships in Ivison et al. (2007).}
\label{fig:offset1}
\end{figure}
Positional errors of radio sources can be ascribed to two independent sources of error, calibration errors and noiselike errors (Condon et al. 1998). Calibration errors are independent of source strength and are best estimated by a comparison with external, more accurate data. In contrast, the noiselike contribution to positional errors is a function of the signal to noise ratio of the detection and is thus the dominant contributor to the positional errors of sources detected at low signal to noise. The noiselike positional errors of radio sources are usually estimated from the models of Condon (1997) for the propagation of errors in 2-dimensional Gaussian fits in the presence of Gaussian noise. However in recent work Ivison et al. (2007) derived a simplified expression for the positional errors due to the Gaussian fits in the special instance of all the radio sources being unresolved by the symmetric beam. In this instance the positional errors can be described by the following equation: \begin{equation}
                                                                                                                                                                                                                                                                                                                                                                                                                                                                                                     \sigma_{\mathrm{fit}}\simeq 0.6\frac{\theta_{\mathrm{N}}}{\mathrm{SNR}}
\label{eq:snr}
                                                                                                                                                                                                                                                                                                                                                                                                                                                                                                    \end{equation} with SNR being the signal to noise ratio of the detected source. We adopt this simplified description of the expected positional errors  in our derivation of the $\sigma_{\mathrm{pos}}$.

The final positional errors should also include an estimate of the calibration error term. As there are no other radio catalogues over this area of better positional accuracy with which to make a comparison,  Bondi et al. (2003) chose to estimate  the calibration error term by comparing the position of the sources in the final mosaiced image with their positions in the images of the single VLA pointings. They find on the basis of this comparison that these calibration errors are of the order of 0.1~arcsec. Thus we adopt as the expected positional error:
 \begin{equation}
  \sigma_{\mathrm{\mathrm{pos}}}^2=0.1^2+\sigma_{\mathrm{fit}}^2
 \label{eq:sum}
 \end{equation}
                                                                                                                                                                                                                                                                                                                                                                                                                                                                                                                                                                                                                                                                                                                                                                                                                                                                                                                                                                                                                                                                                                                                                                                                                                                                                                                                                                                                                                                                                                                                                                                                                                                                                                                                                                                                                                                                                                                                                                                                                                                                                                                                                                                                                                                                                                                                                                                                                                                                                                                                                                                   
 We justify this simplified description of the errors based on figure \ref{fig:offset1} which presents a comparison of these simplified estimates  with those predicted by the method of Condon (1997) and demonstrates that they are in reasonable agreement. Furthermore to account for the possibility that the radio and observed-frame $K_{\mathrm{s}}$ band emission may not arise at exactly the same position in the galaxy we impose the restriction that $\sigma_{\mathrm{pos}}>$~0.5 arcsec.

The $n(m)$ term of the likelihood ratio is estimated from the source counts of the input VIDEO catalogue normalised to the area of the survey.

The most difficult term to estimate in the LR is $q(m)$, the magnitude distribution of the true counterparts to the radio sources. This distribution is estimated using the method outlined in Ciliegi et al. (2003) which begins by calculating the magnitude distribution of all the possible counterparts within a fixed search radius $r_{\mathrm{max}}$ of the radio positions. This distribution is referred to as $\mathrm{total(m)}$. The contribution due to the background source counts is subtracted from $\mathrm{total(m)}$  to produce a magnitude distribution of the excess infrared sources detected around the radio positions, designated as real(m). Thus \begin{equation}
             \mathrm{                                                                                                                                                                                                                                                                                                                                                                                                                                                                                                                                  real}(m)=\mathrm{total}(m)-\left(n(m)*N_{\mathrm{radio}}*\pi*r_{\mathrm{max}}^2\right)
                                                                                                                                                                                                                                                                                                                                                                                                                                                                                                                                                 \end{equation}
 where $N_{\mathrm{radio}}$ is the total number of radio sources in the input catalogue. 

The $q(m)$ distribution is derived from real($m$) by normalising real($m$) and scaling it by a factor $Q_{0}$, where $Q_{0}$ is an estimate of the fraction of radio sources with infrared counterparts above the magnitude limit of the VIDEO survey. Hence:\begin{equation}
                                                                                                                                                                                                                                      q(m)=\frac{\mathrm{real}(m)}{\sum_{m}\mathrm{real}(m)}\times \mathrm{Q}_{0}
                                                                                                                                                                                                                                     \end{equation}

\begin{figure}
 \includegraphics[width=1.0\columnwidth]{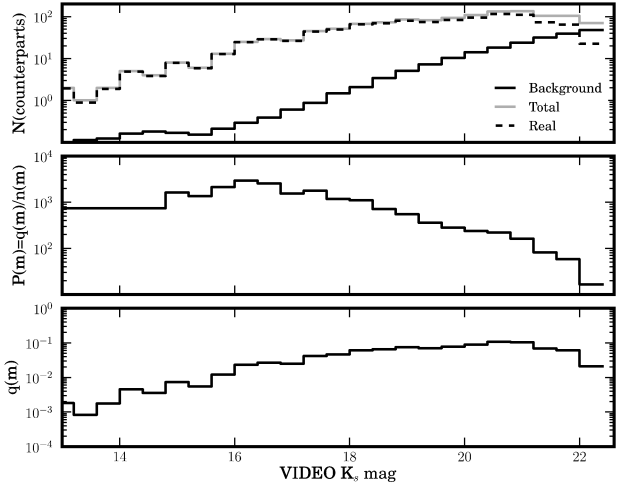}
\caption{The distribution of $\mathrm{real}(m),\mathrm{total}(m)$ and $q(m)$ calculated by the LR in our cross-matching procedure.}
\label{fig:qm}
\end{figure}

The $Q_{0}$ term is usually estimated by determining the fraction of sources with radio counterparts above the background as follows:
\begin{equation}
 \mathrm{Q}_{0}=\frac{N_{\mathrm{matches}}-(\sum_m n(m)\times \pi r_{\mathrm{max}}^2\times N_{\mathrm{radio}})}{N_{\mathrm{radio}}}
 \label{eq:cil}\end{equation}
 where $N_{\mathrm{matches}}$ is the number of possible counterparts within $r_{\mathrm{max}}$ of the radio positions. However in the case of a large $r_{\mathrm{max}}$ we find that this expression leads to an overprediction of the $Q_{0}$ value as a result of a large number of excess sources in the search radii above the predicted background source counts. This effect may result from the tendency for radio-loud AGN to favour denser environments than normal galaxies (Best et al. 2005; Falder et al. 2010), resulting in a large number of close neighbours for these sources. To overcome this difficulty we use the method developed by Fleuren et al. (submitted) to estimate $Q_{0}$. 

This method determines the number of sources in the radio catalogue which have no NIR counterparts within search radius r of the radio positions as a function of the size of the search radius. This function $U_{\mathrm{obs}}(r)$ is compared to $U_{\mathrm{random}}(r)$ which is the number of sources with no NIR counterparts within search radius r when considering an input catalogue with the same number of sources as our original radio catalogue but with sources placed at random positions in the field of view. Fleuren et al. demonstrate that the $U_{\mathrm{obs}}(r)$ and $U_{\mathrm{random}}(r)$ quantites can be related to $Q_{0}$ via the following equation: \begin{equation}
\frac{\mathrm{U}_{\mathrm{obs}}(r)}{\mathrm{U}_{\mathrm{random}}(r)}=1-\mathrm{Q}_{0}F(r)  \label{eq:fleur}\end{equation}  

where: \begin{align}
        F(r)=&\int_{0}^rP(r')dr'=1-e^{\frac{r^2}{2\sigma^2}}\notag \\
        P(r)=&2\pi rf(r) 
       \end{align}

Thus we determine Q$_{0}$ via a fit to the observed ratio of U$_{\mathrm{obs}}(r)$ to  U$_{\mathrm{random}}(r)$. We present a comparison of the Q$_{0}$ estimates produced by the two methods in Table~\ref{tab:q0} and adopt the estimates produced by the Fleuren et al.  method in our  LR matching procedures at all resolutions. 
\begin{table}
\caption{Comparison of $Q_{0}$ estimates using the methods of Ciliegi et al. (2003) and Fleuren et al. (2011).}
\label{tab:q0}
\begin{center}
\begin{tabular}{lllll}
\hline
& & 6~arcsec & 10~arcsec & 15~arsec\\ 
\hline
$K_{s}<22.6$ & $Q_{0}$ (eqn \ref{eq:cil}) & 1.03 & 1.19 & 1.3\\
         & $Q_{0}$ (eqn \ref{eq:fleur}) & 0.90 & 0.90 & 0.90\\
$K_{s}<20.0$ & $Q_{0}$ (eqn \ref{eq:cil}) & 0.51 & 0.55 & 0.57\\
& $Q_{0}$ (eqn \ref{eq:fleur}) & 0.49 & 0.49 & 0.49\\
\hline
\end{tabular}
\end{center}
\end{table}

The presence or absence of more than one infrared counterpart for a particular radio source provides extra information to that contained in the LR itself which can then be used to estimate the reliability of the counterpart source, or the probability that a particular source is the correct counterpart. The reliability is calculated as: 
\begin{equation}
\mathrm{Rel_{i}=\frac{LR_{i}}{\Sigma_{j}LR_{j}+(1-Q_0)}},
\end{equation}
where $j$ is the index of all the possible counterparts to the radio source. We accept sources with Rel$_{i} >$0.8 as being a reliably identified counterpart to the radio source. We are then able to estimate the number of contaminating false identifications $N_{\mathrm{cont}}$ in our catalogue of reliable identifications as being 
\begin{equation}
N_{\mathrm{cont}}=\sum_{\mathrm{Rel}>0.8}(1-\mathrm{Rel}) .                                                                                                                                \end{equation}

\section{Simulated Catalogue}\label{sec:sim}

This study aims to determine whether the LR cross-matching procedure will allow us to reliably identify a large fraction of the counterparts to the radio sources detected in future radio continuum surveys including  EMU, MIGHTEE and WODAN. We aim to compare the performance of this technique at 6 arcsec resolution with those of observations with 10  and 15~arcsec resolution, which are the proposed resolutions of the EMU and WODAN surveys respectively. 
To simulate the degradation of positional accuracy in the Bondi et al. (2003) catalogue which would take place if these observations were performed to the same depth with a larger synthesised beam we add Gaussian scatter to the positions of the radio sources in the catalogue in line with the theoretical predictions of the equations \ref{eq:snr} and \ref{eq:sum} in section \ref{sec:lr}. We generate 100  simulated 'low resolution' catalogues with simulated FWHM beamwidths of 10 and 15 arcsec.  The limitation of this approach is that it precludes us from studying the instances where close pairs of radio sources merge within the larger synthesised beam and the impact this blending may have on our ability to make reliable counterpart identifications.  We attempt to estimate the effect of blended radio sources separately from our LR analysis in section 7.

\section{Near-infrared counterparts to radio sources}

We use the LR technique to find counterparts for both the original VLA radio catalogue and the two sets of 100 simulated catalogues with nominal FWHM beamwidths of 10 and 15 arcsec. We first match the radio sources to almost the full depth of the VIDEO catalogue with $K_{\mathrm{s}}<$~22.6 and then repeat this procedure with the VIDEO catalogue restricted to detections with $K_{\mathrm{s}}<$~20.0.  In order to ensure that our search radius  includes all possible real counterparts to the radio sources  we set r$_{\mathrm{max}}$ to  5 times the largest expected positional error $\sigma_{\mathrm{pos}}$ at each of the three resolutions considered in this study. This results in an r$_{\mathrm{max}}$ of 3.6, 6.0 and 9.0~arcsec in the LR analysis of the 6,10 and 15 arcsec catalogues respectively. The $f(r)$ term of the LR is also adjusted to account for the increased positional uncertainty in the lower resolution catalogues. A summary of the relevant parameters used in the LR analysis at the three different resolutions when matching against the deeper and shallower near-infrared catalogue is given in Tables~\ref{tab:LR} and \ref{tab:LR20}  respectively. For the simulated catalogues these tables contain the mean and standard deviation of the 100 LR matching procedures performed at each resolution. 

\begin{table}

\newcommand{\mc}[3]{\multicolumn{#1}{#2}{#3}}
\caption{Summary of relevant parameters in the LR analysis of the 6, 10 and 15~arcsec catalogues when matching against the VIDEO catalogue with $K_{\mathrm{s}}<$~22.6. N(Rel$>$0.8) and \% Rel$>$0.8 represents the number and percentage of radio sources  which have  counterparts  with Rel$>$0.8. The total number of radio sources in the input catalogue is 1031. Similarly N$_{\mathrm{cont}}$ and \% cont represent estimates of the number and percentage contribution of misidentified contaminating sources. }\label{tab:LR}
\begin{tabular}{lllll}
\hline
 &  & 6 arcsec & 10 arcsec& 15 arcsec\\\hline
\mc{1}{l}{$r_{\mathrm{max}}$}[arcsec] & × & 3.6& 6.0& 9.0\\
\mc{1}{l}{$Q_0$} & × & 0.90 & 0.90& 0.90\\
\mc{1}{l}{$N$(Rel$>$0.8)} & × & 915 & 887.32 $\pm$ 5.91& 837.91~$\pm$~7.18\\
\mc{1}{l}{$N_{\mathrm{no\;match}}$}& & 68& 50.9 $\pm$ 1.71& 32.63~$\pm$~1.75\\
\mc{1}{l}{$N<$ r$_{\mathrm{max}}$} & × & 1274 & 1809.34 $\pm$9.36 & 2669.30~$\pm$~12.61\\
\mc{1}{l}{$N_{\mathrm{cont}}$} & × & 6.825 & 12.405 $\pm$0.809 & 19.346~$\pm$~0.93\\
\mc{1}{l}{\% Rel$>$0.8}& × & 88.7\%  & 86.0 $\pm$ 0.6\% & 81.8 $\pm$ 0.7\%\\
\mc{1}{l}{\% cont}& & 0.74\% & 1.4 $\pm$ 0.1\%& 2.3 $\pm$ 0.1\%\\
\hline
\end{tabular}
\end{table}

\begin{table}
\caption{Summary of relevant parameters in the LR analysis of the 6, 10 and 15~arcsec catalogues with $K_{\mathrm{s}}<$~20.0. The row headings are as in Table~\ref{tab:LR}.}
\label{tab:LR20}
\newcommand{\mc}[3]{\multicolumn{#1}{#2}{#3}}
\begin{tabular}{lllll}
\hline
× & × & 6 arcsec & 10 arcsec& 15 arcsec\\\hline
\mc{1}{l}{$r_{\mathrm{max}}$}[arcsec] & × & 3.6& 6.0& 9.0\\
\mc{1}{l}{$Q_0$} & × & 0.49 & 0.49& 0.49\\

\mc{1}{l}{$N$(Rel$>$0.8)} & × & 486 & 490.31 $\pm$ 3.65& 485.09~$\pm$~5.08\\
\mc{1}{l}{$N_{\mathrm{no\;match}}$}& & 510& 484.01 $\pm$ 2.22& 437.53~$\pm$~3.30\\
\mc{1}{l}{$N<$ r$_{\mathrm{max}}$} & × & 567 & 645.91 $\pm$3.02 & 775.81~$\pm$~5.19\\
\mc{1}{l}{$N_{\mathrm{cont}}$} & × & 3.931 & 6.839 $\pm$0.555 & 11.062~$\pm$~0.708\\
\mc{1}{l}{\% Rel$>$0.8}& × & 47.1\%  & 47.5 $\pm$ 0.6\% & 47.0 $\pm$ 0.5\%\\
\mc{1}{l}{\% cont}& & 0.80\% & 1.4 $\pm$ 0.1\%& 2.3 $\pm$ 0.1\%\\
\hline
\end{tabular}
\end{table}

\begin{figure}
 \includegraphics[width=1\columnwidth]{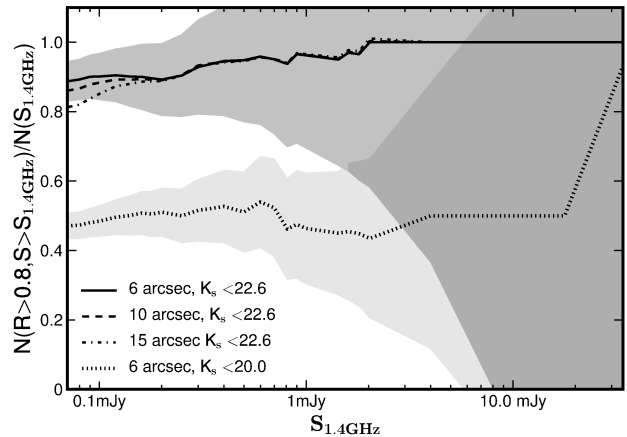}
\caption{The fraction of reliable counterparts detected at 6, 10 and 15~arcsec resolution when matching against the VIDEO NIR catalogue restricted to detections with $K_{\mathrm{s}}<$~22.6 and  $K_{\mathrm{s}}<$~20.0. The greyscale bands represent the 1$\sigma$ Poisson error on the cross-matched fractions.}
\label{fig:complete}
\end{figure}

\begin{figure}
 \includegraphics[width=1\columnwidth]{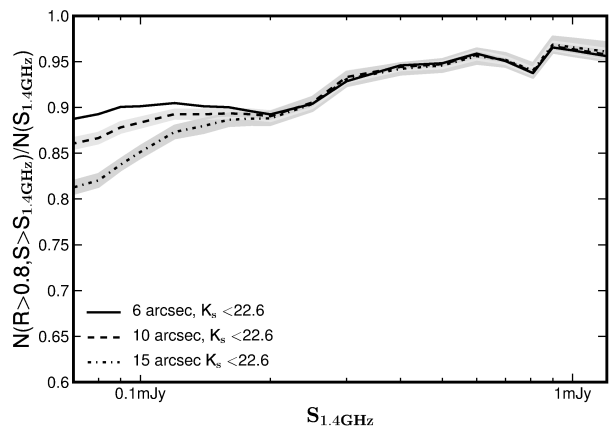}
\caption{Close-in plot of the fraction of reliable counterparts detected for the faint radio sources ($<$~1~mJy) at 6,10 and 15~arcsec resolution when matching against the VIDEO NIR catalogue restricted to detections with $K_{\mathrm{s}}<$~22.6. The greyscale filled bands represent the 1$\sigma$ variation between the 100 simulated low resolution radio catalogues and do not include the Poisson errors.}
\label{fig:completezoom}
\end{figure}

\subsection{Counterparts as a function of resolution}

An inspection of Table~\ref{tab:LR} reveals that when cross-matching against the full VIDEO catalogue the number of sources with reliable counterparts decreases with decreasing resolution. We identify 915, 887 and 838, sources with reliable counterparts at 6,10 and 15~arcsec resolution, with each decrease in resolution resulting in a loss of approximately 3\%  and then a further 4\% of the identifications. The completeness as a function of flux density for all three resolutions is plotted in figures \ref{fig:complete} and \ref{fig:completezoom}. These figures indicate that the number of lost identifications  increase at  lower flux densities where the lower signal to noise ratio of the detections result in larger positional uncertainties. For clarity figure \ref{fig:completezoom} presents a close-in view of the completeness at the fainter flux densities ($<$~1mJy) and the greyscale filled regions in this figure indicate the $1\sigma$ variation between our 100 simulated catalogues at each resolution.

Encouragingly, our results indicate that at 6~arcsec resolution we are able to identify nearly all the available counterparts whose magnitudes are less than the imposed $K_{\mathrm{s}}<$~22.6 magnitude limit as the fraction of reliably identified sources 89\% is very close to our estimated Q$_0$ value of 0.90. Our estimates also indicate that the contribution of contaminating or misidentified sources is very low at $\sim$ 0.7\%.  Table~\ref{tab:LR} also reveals that our estimate of the number of contaminating sources in our cross-matched catalogue increases at lower resolution to 1.4\% and 2.3\% at resolutions of 10 and 15~arcsecs.

A subsection of the radio data in this paper has previously been matched to deep $K-$band data using a LR procedure (Ciliegi et al. 2005). This matching was performed to a $K-$band depth of 23.9 over a 165 arcmin$^2$ field observed by Iovino et al. (2005), the limiting magnitude used in the matching procedure corresponds to the 50~per~cent completeness limit of the Iovino survey. Ciliegi et al. 2005 find a total of 43 reliable K-band matches to the 65 radio sources located within this subfield, corresponding to a completeness of $\sim$ 66~per~cent which is significantly lower than the 88.7~per~cent completeness achieved in this work. We ascribe this improvement to the greater depth of the VIDEO survey, which has factor of $\sim$12 greater integration time over the Iovino catalogue with a telescope of similar aperture. Furthermore the $q(m)$ distributions and LR in Ciliegi et al. (2005) are derived from the VVDS optical catalogues (McCracken et al. 2003) available over the whole 1 square degree radio field. As $q(m_{opt})$ distributions are not precisely equivalent to $q(m_{NIR})$ it is likely that their use of the optical magnitude distribution in the matching procedure contributes to an underestimate of the significance of some of the fainter NIR matches.

\subsection{Counterparts as a function of near-infrared magnitude}

In the case of matching against the VIDEO catalogue limited to the depth of the VHS, Table~\ref{tab:LR20} reveals a similar increasing trend in the number of contaminating sources with decreasing resolution from 0.8~per~cent at 6~arcsec to 1.4 and 2.3~per~cent at the lower resolutions. However the completeness of the cross-matched catalogue is nearly identical at all three resolutions, indicating that the depth of the complementary near-infrared data is a more relevant limiting factor at these shallower survey depths than radio survey resolution. This trend can be understood by examining the middle plot in figure \ref{fig:qm}, which indicates that NIR counterparts with magnitudes lower than $K_{\mathrm{s}}<$~20.0  are assigned higher $q(m)/n(m)$ fractions than fainter NIR matches. The intrinsic rarity of brighter NIR sources thus increases the significance of these bright NIR matches allowing us to partially overcome the limitation of poorer positional accuracy. In contrast at deeper NIR magnitudes the increasing density of faint sources dictates that resolution, or equivalently positional accuracy, is increasingly relevant in determining the correct counterpart. 

\begin{figure}
 \includegraphics[width=1.0\columnwidth]{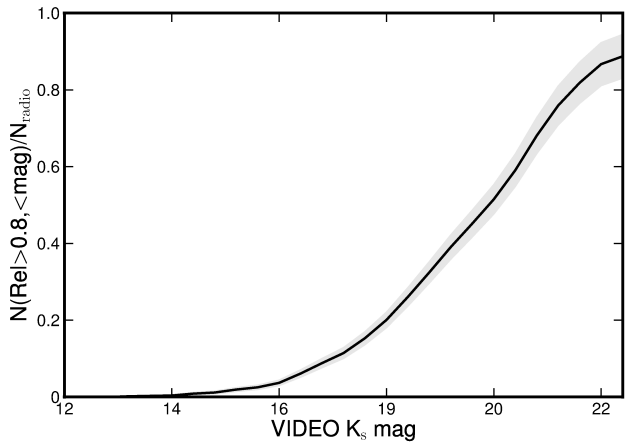}
\caption{The fraction of reliably associated counterparts as a function of $K_{\mathrm{s}}$ magnitude at 6~arcsec resolution for the depth of our radio imaging data. The greyscale bands represent the 1$\sigma$ Poisson error on the cross-matched fraction.}
\label{fig:magk}
\end{figure}

In figure \ref{fig:magk}, we show a plot of the fraction of reliably identified counterparts as a function of $K_{\mathrm{s}}$ band magnitude. This demostrates that deep near-infrared and/or optical data are crucial for successfully identifying faint radio sources, at least to the depth of our radio imaging data and this will only become more of an issue for yet deeper radio imaging such as those planned with the MeerKAT (e.g. Jarvis 2011). We note that once again the reliably identified fraction of radio sources is very close to the $Q_{0}$ estimate indicating that we are identifying nearly all the available counterparts with the NIR catalogue to $K_{\rm s} <22.6$. 

\subsection{Mis-identified counterparts at low resolution}

Apart from considering changes to the overall completeness with resolution it is also of interest to determine whether there are differences between the low and high resolution catalogues in terms of the subset of radio sources that have reliably identified counterparts and whether these radio sources  are associated with the same NIR counterpart in all cases.  Changes in the exact composition of the output cross-matched catalogues occur because the Gaussian scatter introduced to the positions of the simulated low resolution radio sources will alter their relative position to any possible near-infrared counterparts, furthermore the lower resolution catalogues have larger positional uncertainties $\sigma_{\mathrm{pos}}$. These two factors result in changes to the $f(r)$ term of the LR and consequently alter the overall statistical significance of a match between any pair of sources. Changes in the cross-matched low resolution catalogues compared to the original VLA cross-matched catalogue occur in three different forms, radio sources with secure identifications at 6~arcsec no longer have secure identifications in the low resolution catalogue, we refer to these as R$_{\mathrm{lose}}$, radio sources with no reliable counterparts at 6~arcsec have an identified counterpart at lower resolution R$_{\mathrm{gain}}$.  Radio sources with secure identifications at both resolutions R$_{\mathrm{common}}$  are identified with a different NIR counterpart at different resolutions R$_{\mathrm{diff\: id}}$. The average changes in the composition of the output catalogues are listed in Tables~\ref{dif1} and \ref{tab:dif2}, these are relative to the original VLA cross-matched catalogue in all cases. These tables indicate that differences in the exact composition of the output cross-matched catalogues at the three different resolutions are usually small. When matching against the deeper VIDEO catalogue 95 and 88~per~cent of the radio sources in the original cross-matched catalogue are identified with the identical NIR counterpart in the 10 and 15 arcsec catalogues. The catalogue with shallower NIR magnitude limits produced even fewer discrepancies with 98 and 95~per~cent of the identifications remaining unchanged at lower resolution. This clearly illustrates the increased difficulty in associating the radio sources with their correct counterparts when matching against very deep complementary datasets. 

\begin{table}
\caption{Summary of differences between the cross-matched catalogues created when matching against the original 6~arcsec resolution catalogue and the simulated low resolution catalogues at 10 and 15 arcsec. This table summarises the differences when matching against the VIDEO NIR catalogue limited to $K_{\mathrm{s}} <$~22.6.}
\label{dif1}
\newcommand{\mc}[3]{\multicolumn{#1}{#2}{#3}}
\begin{center}
\begin{tabular}{llll}
\hline
× & ×  & 10 arcsec& 15 arcsec\\\hline
\mc{1}{l}{$R_{\mathrm{lose}}$} & × & 41.80~$\pm$~4.78& 94.10~$\pm$~6.36\\
\mc{1}{l}{$R_{\mathrm{gain}}$}& & 14.12 $\pm$ 2.89& 17.00~$\pm$~3.02\\
\mc{1}{l}{$R_{\mathrm{common}}$} & ×  & 873.20 $\pm$4.78 & 820.91~$\pm$~6.36\\
\mc{1}{l}{$R_{\mathrm{diff\: id}}$} & × & 5.10 $\pm$2.59 & 9.97~$\pm$3.22\\
\hline
\end{tabular}
\end{center}
\end{table}

\begin{table}
\newcommand{\mc}[3]{\multicolumn{#1}{#2}{#3}}
\caption{Summary of differences between the cross-matched catalogues created when matching against the original 6~arcsec resolution catalogue and the simulated low resolution catalogues at 10 and 15 arcsec resolution. This table summarises the differences when matching against the VIDEO NIR catalogue limited to $K_{\mathrm{s}}<$~20.0.} 
\label{tab:dif2} 
\begin{center}
\begin{tabular}{llll}
\hline
× & ×  & 10 arcsec& 15 arcsec\\\hline
\mc{1}{l}{$R_{\mathrm{lose}}$} & × & 9.71~$\pm$~2.69& 23.62~$\pm$~3.86\\
\mc{1}{l}{$R_{\mathrm{gain}}$}& & 14.02 $\pm$ 2.42& 22.72~$\pm$~2.86\\
\mc{1}{l}{$R_{\mathrm{common}}$} & ×  & 476.30 $\pm$3 & 462.37~$\pm$~3.85\\
\mc{1}{l}{$R_{\mathrm{diff\: id}}$} & × & 0.62 $\pm$0.82 & 2.02~$\pm$1.36\\
\hline
\end{tabular}
\end{center}
 \end{table}

\subsection{Redshift distributions of identified radio sources}

\begin{figure}
 \includegraphics[width=1\columnwidth]{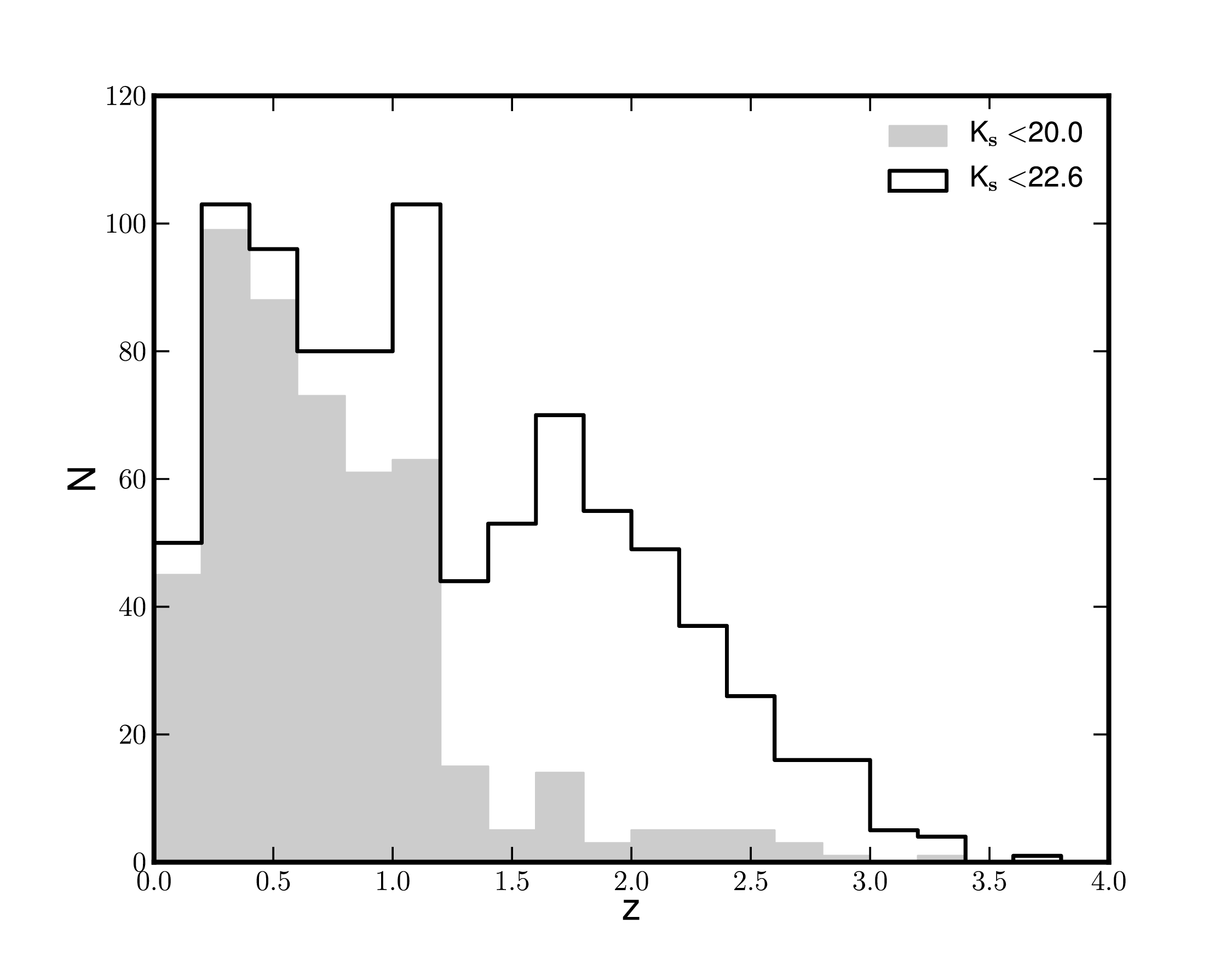}
\caption{The photometric redshift distribution of the counterparts in the matched catalogues restricted to sources with magnitudes $K_{\mathrm{s}}<$~22.6 and $K_{\mathrm{s}}<$~20.0. }
\label{fig:histo}
\end{figure}

Photometric redshifts for the combined VIDEO and CFHTLS-D1 datasets have been derived using SED fitting techniques with the photometric redshift package Le Phare (Arnouts et al. 1999; Ilbert et al. 2006), based on simulations these redshifts have accuracies of $\sigma_z \sim$0.1 for sources with $K_{\mathbf{s}}<$~22.6, and for a small sample of real objects with spectroscopic redshifts from the VVDS survey (Le F{\`e}vre et al. 2007) $\sigma_z \sim$0.095. Figure \ref{fig:histo} presents the redshift distribution of sources matched at 6~arcsec to almost the full depth of the VIDEO survey with those restricted to matches with $K_{\mathbf{s}}<$~20.0.  Unsurprisingly there is clear evidence for a decline in the fraction of high redshift sources detected in the survey with shallower magnitude limits. It is clear that while the VHS survey will allow us to identify a significant fraction,$\sim$ 83~per~cent, of the sources with z$<$~1.2  a large fraction of the radio sources at redshifts higher than this threshold will not be present in this shallower wide-field survey. The histogram indicates that only 14~per~cent of the identified counterparts at z~$>$~1.2 had magnitudes brighter than the $K_{\mathbf{s}}<$~20.0 limit. There is also evidence for a loss of a significant ($\sim$40~per~cent) number of counterparts in the 1$\lesssim$~z$\lesssim$~1.2 redshift bin. 

\section{Blended Sources}

As mentioned in section \ref{sec:sim}, our LR investigation makes no attempt to consider the possible effects of an increased number of blended sources at lower resolutions on the completeness produced by the LR technique. To estimate the increase in the number of radio sources which will be blended by the beam in 10 and 15 arcsec images, compared to the 6 arcsec resolution image, we use the size and spatial distribution of radio sources predicted in the Square Kilometer Array Simulated Skies (Wilman et al. 2008; 2010).  From this simulation we extract a list of radio source components with $S_{\mathrm{1.4GHz}}$ greater than the predicted  5$\sigma$ flux density limit of the EMU survey of 50~$\mu$Jy over a 1~square degree field of view. The size of the radio components are adjusted to simulate the effect of convolution with a Gaussian beam of FWHM size of 6, 10 and 15 arcsec respectively.

 Separating close components in an astronomical image is a complex problem and source extraction packages adopt a variety of approaches to this task. For instance the \textsc{aips} SAD task attempts to separate emission features into multiple components based on the level of residual flux present after a single component Gaussian fit, whereas \textsc{sextractor}  maps the detected emission at a number of sub-thresholds to detect junctions in the emission profile of the blended source (Bertin \& Arnouts 1996). Consequently the probability of separating a pair of close sources is difficult to quantify and depends on the characteristics of the pair  including  their separation, relative angular sizes and peak fluxes as well as the details of the deblending technique in question. To obtain an estimate of the fraction of blended pairs in our catalogues we make the simplifying assumption that a source extraction algorithm will be unable to separate a pair of sources if their separation is less than the mean of their FWHM. In this determination we disregard the contribution of very extended sources ($>$~17 arcsec) as these should be fully resolved by the beam, these extended sources constitute less than 5~per~cent of the total source population at our chosen flux density limits. 

To confirm that this simplification is reasonable we created maps of the simulated radio source components at 6, 10 and 15 arcsec resolution  using the Simulated Skies S3Map tool (Levrier et al. 2009). The \textsc{aips} SAD task was used to extract a source list from these images and an inspection of the output component lists confirmed that the the pairs selected using our separation criteria were detected as a single Gaussian component by the SAD algorithm.  Based on our separation criteria and an inspection of the SAD outputs we determine the number of detected radio sources in the simulated field which consist of a blend of two or more underlying radio source components. We consider separately detected radio sources which consist of a blend of multiple radio components  arising from the same radio galaxy (i.e. radio lobes from an FRI/FRII source blended together) and sources which are a blend of unrelated radio galaxies, the results of our analysis are presented in Table~\ref{table:blend}.

\begin{table}

\newcommand{\mc}[3]{\multicolumn{#1}{#2}{#3}}
\caption{The number of blended sources detected at 6, 10 and 15~arcsec resolution predicted by the SKA Simulated Skies (Wilman et al. 2008; 2010). Column 1 indicates the number of components blended per detected source.}  
\label{table:blend} 
\begin{tabular}{llll}
\hline
\mc{1}{c}{Number of components} & 6~arcsec & 10~arcsec& 15~arcsec\\\hline
& \mc{3}{c}{unrelated blends}\\
\hline
\mc{1}{c}{×} & × & ×\\
{2} & 22 & 36 & 62\\\hline
& \multicolumn{3}{c}{multiple component blends}\\
\hline
\mc{1}{c}{×} & × & ×\\
{2} & 218 & 218& 219\\
{3} & 41& 41& 41\\
{4 or more} & 2 & 4& 4 \\
\hline
\end{tabular}
\end{table}

The initial simulated component list  consisted of 1908 components with  $S_{\mathrm{1.4GHz}}>$~50~$\mu$Jy, these are reduced to 1579, 1557 and 1531  detected sources of which 283, 299 and 325 detections are blended sources in the 6, 10 and 15 arcsec catalogues respectively. As the LR is not designed to account for the possibility of two or more counterparts per radio source it is reasonable to assume that sources which consist of a blend of unrelated radio galaxies will not be reliably associated with an appropriate counterpart. The total contribution of these unrelated blends increases from  1.4~per~cent at 6~arcsec resolution to 2.3 and 4.0~per~cent at 10 and 15~arcsec resolutions respectively. Thus we conclude that the contribution of this effect to incompleteness in any LR based cross-matching routine in future surveys will be small,  at the level of approximately 1$\sim$2.5~per~cent. The real contribution of this effect will depend on the baseline distribution and uv coverage of the survey in question.  We do not attempt to estimate to what extent the blended multiple component radio sources  will contribute towards incompleteness when cross-matching as the symmetric nature of the SKADS simulation radio sources will not allow us to realistically determine to what extent the positions of the final blended radio sources will deviate from the expected position of the NIR counterpart. 

\section{Towards deeper surveys with MeerKAT}

Although in this paper we have concentrated on our ability to cross-match radio continuum sources from surveys such as those proposed for ASKAP and APERTIF, it is clear that similar techniques may also be appropriate for much deeper ($\sim 100$~nJy) and narrower surveys such as the MeerKAT International Giga-Hertz Tiered Extragalactic Exploration (MIGHTEE) Survey (Jarvis 2011). Although we currently do not have the necessary data to test how well we can recover counterparts to the radio sources at these depths, it is clear that our ability to identify shorter wavelength counterparts becomes worse towards fainter flux densities (e.g. Figure~\ref{fig:completezoom}) due to both the declining signal-to-noise and increasing density of counterpart sources. However resolution of the radio maps is also a contributing factor. The issue of spatial resolution will only become more difficult to deal with at deeper flux densities due to the edging closer to the classical confusion level of  $\sim$25 beams per source(Condon 1974). The current design of MeerKAT, incorporating 20~km maximum baselines, will provide a spatial resolution of $\sim 3$~arcsec at 1.4~GHz which for a 10$\mu$Jy flux-density limit corresponds to $\sim 40$~beams per source according to the simulated skies of Wilman et al. (2010). Thus as we approach the new parameter space in flux-density and survey area offered by MeerKAT, and eventually the SKA, the crucial aspect of telescope design will be how to achieve $\sim 1$~arcsec resolution coupled with high surface brightness sensitivity.

\section{Conclusions}

We have presented a comparison of the infrared counterparts identified by the LR technique to radio sources observed with synthesized beamwidths of  6, 10 and 15~arcsec resolution when matched against a NIR catalogue at depths of $K_{\mathbf{s}}<$~22.6 and 20.0. The results of our analysis indicate that we are able to reliably associate nearly all the available radio source counterparts in the NIR catalogue, limited to $K_{\mathrm{s}}<$~22.6, with the appropriate radio source. Furthermore $\sim$93 and 88~per~cent of the identifications made by this technique remain unchanged  when matching the lower resolution 10 and 15~arcsec catalogues to the deeper NIR catalogue. At all resolutions the technique delivers a catalogue with a high degree of completeness and a low percentage of contaminating misidentified sources. When matching against the shallower NIR catalogue the fraction of unchanged counterparts increases to 97 and 94~per~cent in the 10 and 15~arcsec cases. 

Although changes in completeness and contamination fractions in our study were all relatively small as a function of resolution it is clear that both of these quality indicators degrade systematically as the depth of the matching complementary data increases and the resolution of the radio data decreases.  

We conducted a brief investigation into the question of unrelated radio sources being blended together at lower resolutions and reducing the completeness achieved by the LR cross-matching technique. We conclude that at a flux density limits of $\sim$50~$\mu$Jy, comparable to the EMU wide-field survey,  the contribution of this effect is likely to be small at the level of only a few percent. Finally the distribution of photometric redshifts in our matched catalogues indicates that a significant fraction of the radio sources at z$\gtrsim$1.2 will not be detected in the shallower wide-field VISTA Hemisphere survey.

\section*{ACKNOWLEDGEMENTS} 
Kim McAlpine acknowledges the support of a South African SKA Bursary. MJJ acknowledges the support of an RCUK fellowship. We would also like to thank Marco Bondi for providing us with the VLA data. We also thank Loretta Dunne and Ray Norris for useful discussions about cross-matching in general.


\begin{thebibliography}{}

\bibitem[\protect\citeauthoryear{{Arnouts}, {Cristiani}, {Moscardini},
  {Matarrese}, {Lucchin}, {Fontana} \& {Giallongo}}{{Arnouts}
  et~al.}{1999}]{Arnouts}
{Arnouts} S.,  {Cristiani} S.,  {Moscardini} L.,  {Matarrese} S.,  {Lucchin}
  F.,  {Fontana} A.,    {Giallongo} E.,  1999, MNRAS, 310, 540

\bibitem{Baldry}{Baldry} I.~K. et al., 2010, MNRAS, 404, 86

\bibitem[\protect\citeauthoryear{{Bertin} \& {Arnouts}}{{Bertin} \&
  {Arnouts}}{1996}]{Bertin}
{Bertin} E.,  {Arnouts} S.,  1996, A\&AS, 117, 393

\bibitem{Best2005}{Best} P.~N.,  {Kauffmann} G.,  {Heckman} T.~M.,  {Brinchmann} J.,  {Charlot}
  S.,  {Ivezi{\'c}} {\v Z}.,    {White} S.~D.~M.,  2005, MNRAS, 362, 25

\bibitem{Bondi2003}Bondi M. et al., 2003, A\&A, 403, 857

\bibitem[\protect\citeauthoryear{{Ciliegi}, {Zamorani}, {Hasinger}, {Lehmann},
  {Szokoly} \& {Wilson}}{{Ciliegi} et~al.}{2003}]{Cil2003}
{Ciliegi} P.,  {Zamorani} G.,  {Hasinger} G.,  {Lehmann} I.,  {Szokoly} G.,
  {Wilson} G.,  2003, A\&A, 398, 901

\bibitem{Cil2005}{Ciliegi}, P. et al., 2005, A\&A, 441,879
\bibitem{Condon1974}{Condon} J.~J., 1974, ApJ, 188, 279

\bibitem[\protect\citeauthoryear{{Condon}}{{Condon}}{1997}]{Condon1997}
{Condon} J.~J.,  1997, PASP, 109, 166

\bibitem[\protect\citeauthoryear{{Condon}, {Cotton}, {Greisen}, {Yin},
  {Perley}, {Taylor} \& {Broderick}}{{Condon} et~al.}{1998}]{Condon1998}
{Condon} J.~J.,  {Cotton} W.~D.,  {Greisen} E.~W.,  {Yin} Q.~F.,  {Perley}
  R.~A.,  {Taylor} G.~B.,    {Broderick} J.~J.,  1998, AJ, 115, 1693

\bibitem[\protect\citeauthoryear{{de Ruiter}, {Arp} \& {Willis}}{{de Ruiter}
  et~al.}{1977}]{DeRuit}
{de Ruiter} H.~R.,  {Arp} H.~C.,    {Willis} A.~G.,  1977, A\&AS, 28, 211

\bibitem[Falder et al.(2010)]{2010MNRAS.405..347F} Falder, J.~T., Stevens, 
J.~A., Jarvis, M.~J., et al.\ 2010, MNRAS, 405, 347 


\bibitem{greis}{Greisen} E.W., 2003, in Heck A.,ed.,Information 
Handling in Astronomy -- Historical Vistas, Kluwer 
Academic Publishers, Dordrecht, p. 109

\bibitem{Ilbert}Ilbert O. et al., 2006, A\&A, 457, 841

\bibitem{Iovino}{Iovino}, A. et al.,2005, A\&A, 442, 423

\bibitem{Ivison}{Ivison} R.~J. et al., 2007, MNRAS, 380, 199

\bibitem[\protect\citeauthoryear{{Jarvis}}{{Jarvis}}{2011}]{Jarvis2011}
{Jarvis} M.~J.,  2011, preprint, (astro-ph/1107.5165)

\bibitem[\protect\citeauthoryear{{Jarvis} \& {Rawlings}}{{Jarvis} \&
  {Rawlings}}{2004}]{JR2004}
{Jarvis} M.~J.,  {Rawlings} S.,  2004, NewAR, 48, 1173

\bibitem[\protect\citeauthoryear{{Kellermann}, {Fomalont}, {Mainieri},
  {Padovani}, {Rosati}, {Shaver}, {Tozzi} \& {Miller}}{{Kellermann}
  et~al.}{2008}]{Kell2008}
{Kellermann} K.~I.,  {Fomalont} E.~B.,  {Mainieri} V.,  {Padovani} P.,
  {Rosati} P.,  {Shaver} P.,  {Tozzi} P.,    {Miller} N.,  2008, ApJS, 179, 71

\bibitem{lef}{Le F{\`e}vre} O. et al., 2007, A\&A, 439, 845

\bibitem[\protect\citeauthoryear{{Levrier}, {Wilman}, {Obreschkow},
  {Kloeckner}, {Heywood} \& {Rawlings}}{{Levrier} et~al.}{2009}]{s3map}
{Levrier} F.,  {Wilman} R.~J.,  {Obreschkow} D.,  {Kloeckner} H.~R.,  {Heywood}
  I.~H.,    {Rawlings} S.,  2009, in Proceedings of Wide Field Astronomy \&
  Technology for the Square Kilometre Array (SKADS 2009)


\bibitem {McCracken}{McCracken}, H. J. et al., 2003, A\&A, 410, 17-32

\bibitem[Muxlow(2010)]{2010evn..confE..27M} Muxlow, T.\ 2010, ''Proceedings 
of the 10th European VLBI Network Symposium and EVN Users Meeting: VLBI and 
the new generation of radio arrays.~September 20-24, 2010.~Manchester, 
UK.

\bibitem[Myers et al.(2010)]{2010AAS...21535703M} Myers, S.~T., 
construction, E., 
\& commisioning team 2010, Bulletin of the American Astronomical Society, 42, \#357.03 



\bibitem[Norris et al.(2011)]{Norris2011} Norris, R.~P., Hopkins, 
A.~M., Afonso, J., et al.\ 2011, PASA, 28, 215 


\bibitem[\protect\citeauthoryear{{Padovani}, {Mainieri}, {Tozzi}, {Kellermann},
  {Fomalont}, {Miller}, {Rosati} \& {Shaver}}{{Padovani}
  et~al.}{2009}]{Pad2009}
{Padovani} P.,  {Mainieri} V.,  {Tozzi} P.,  {Kellermann} K.~I.,  {Fomalont}
  E.~B.,  {Miller} N.,  {Rosati} P.,    {Shaver} P.,  2009, ApJ, 694, 235

\bibitem[Padovani et al.(2011)]{2011ApJ...740...20P} Padovani, P., Miller, 
N., Kellermann, K.~I., et al.\ 2011, ApJ, 740, 20 



\bibitem[Raccanelli et al.(2011)]{2011arXiv1108.0930R} Raccanelli, A., 
Zhao, G.-B., Bacon, D.~J., et al.\ 2011, arXiv:1108.0930 


\bibitem{Richter}{Richter} G.~A.,  1975, Astronomische Nachrichten, 296, 65
\bibitem{Rott2011}{Rottgering} H. et al., 2011, preprint, (astro-ph/1107.1606)
\bibitem{Sad2002}{Sadler} E.~M. et al., 2002, MNRAS, 329, 227
\bibitem[\protect\citeauthoryear{{Seymour}, {Dwelly}, {Moss}, {McHardy},
  {Zoghbi}, {Rieke}, {Page}, {Hopkins} \& {Loaring}}{{Seymour}
  et~al.}{2008}]{Seymour2008}
{Seymour} N.,  {Dwelly} T.,  {Moss} D.,  {McHardy} I.,  {Zoghbi} A.,  {Rieke}
  G.,  {Page} M.,  {Hopkins} A.,    {Loaring} N.,  2008, MNRAS, 386, 1695
\bibitem[\protect\citeauthoryear{{Simpson}, {Mart{\'{\i}}nez-Sansigre},
  {Rawlings}, {Ivison}, {Akiyama}, {Sekiguchi}, {Takata}, {Ueda} \&
  {Watson}}{{Simpson} et~al.}{2006}]{Simp2006}
{Simpson} C.,  {Mart{\'{\i}}nez-Sansigre} A.,  {Rawlings} S.,  {Ivison} R.,
  {Akiyama} M.,  {Sekiguchi} K.,  {Takata} T.,  {Ueda} Y.,    {Watson} M.,
  2006, MNRAS, 372, 741
\bibitem{Smith}{Smith} D.~J.~B. et al., 2011, MNRAS, 416, 857
\bibitem{Suth}{Sutherland} W.,  {Saunders} W.,  1992, MNRAS, 259, 413
\bibitem{Wilman2008}Wilman R.~J. et al., 2008, MNRAS, 388, 1335
\bibitem[\protect\citeauthoryear{{Wilman}, {Jarvis}, {Mauch}, {Rawlings} \&
  {Hickey}}{{Wilman} et~al.}{2010}]{Wilman2010}
{Wilman} R.~J.,  {Jarvis} M.~J.,  {Mauch} T.,  {Rawlings} S.,    {Hickey} S.,
  2010, MNRAS, 405, 447

\end{thebibliography}
\end{document}